\documentclass[journal=jpcbfk,articletitle=true,manuscript=article]{achemso}
  
\usepackage{graphicx}
\usepackage{amsmath}
\usepackage{amssymb}
\usepackage{dcolumn}
\usepackage{color}
\usepackage{natbib}
\usepackage{ctable}
\usepackage[sort&compress]{natbib}

\title{Hydration free energies of polypeptides from popular implicit solvent models versus all-atom simulation results based on molecular quasichemical theory}

\author{Rohan S. Adhikari}
\affiliation{Department of Chemical and Biomolecular Engineering, Rice University, Houston, TX 77005}
\author{Arjun Valiya Parambathu}
\affiliation{Department of Chemical and Biomolecular Engineering, Rice University, Houston, TX 77005}
\author{Walter G. Chapman}
\affiliation{Department of Chemical and Biomolecular Engineering, Rice University, Houston, TX 77005}
\author{Dilipkumar N. Asthagiri}
\email{dna6@rice.edu}
\affiliation{Department of Chemical and Biomolecular Engineering, Rice University, Houston, TX 77005}

\begin{document}

\begin{abstract}
The hydration free energy of a macromolecule is the central property of interest for understanding its distribution over conformations and its state of aggregation. Calculating the hydration free energy of a macromolecule in all-atom simulations has long remained a challenge, necessitating the use of models wherein the effect of the solvent is captured without explicit account of solvent degrees of freedom. This situation has changed with developments in the molecular quasi-chemical theory (QCT), an approach that enables calculation of the hydration free energy of macromolecules within all-atom simulations at the same resolution as is possible for small molecule solutes. 
The theory also provides a rigorous and physically transparent framework to conceptualize and model interactions in molecular solutions, and thus provides a convenient framework to 
investigate the assumptions in implicit-solvent models. In this study, we compare the results using molecular QCT versus predictions from EEF1, ABSINTH, and GB/SA implicit-solvent models for poly-glycine and poly-alanine solutes covering
a range of chain lengths and conformations.  Among the three models, GB/SA does best in capturing the broad trends in hydration free energy. We trace the deficiencies of the group-additive EEF1 and ABSINTH models to their under-appreciation of the cooperativity of hydration between solute groups; seen in this light, the better performance of GB/SA can be attributed to its treatment of the collective properties of hydration, albeit within a continuum dielectric framework. We highlight the importance of validating the individual physical components that enter implicit solvent models for protein solution thermodynamics. 
\end{abstract}

\begin{tocentry}
\includegraphics{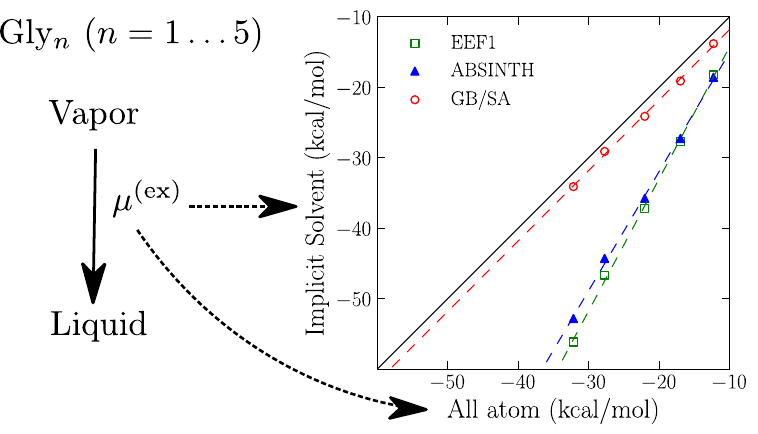}
\begin{center}
       
\end{center}
\end{tocentry}

\section{Introduction}

The structure, function, and phase behavior of biological macromolecules intimately depends on their interaction with the surrounding aqueous solvent matrix \cite{ball:cr2008}.  Formally, the excess chemical potential, $\mu^{\mathrm{(ex)}}$, of the macromolecule in the solvent contains all the information that is relevant to understanding the (thermodynamic) role of the solvent on the solute. $\mu^{\mathrm{(ex)}}$ is given by \cite{widom:jpc82,lrp:book}
\begin{eqnarray}
    \mu^{\mathrm{(ex)}} = k_{\rm B}T \ln \int e^{\varepsilon / k_{\rm B}T} P(\varepsilon) d\varepsilon \, ,
\label{eq:qct}
\end{eqnarray}
where $P(\varepsilon)$ is the probability density distribution of the binding energy ($\varepsilon$) of the solute with the solvent, 
$T$ is the temperature, and $k_{\rm B}$ is the Boltzmann constant.  Understanding and predicting $\mu^{\mathrm{(ex)}}$ for macromolecules remains of long-standing concern in biology and aqueous phase chemistry.

 {\emph{Atomistic}} models treat the solute and the solvent in full atomic detail and the thermodynamics of hydration is in principle, obtainable using Eq.~\ref{eq:qct}. However, for much of the history of computer simulation, applying Eq.~\ref{eq:qct} to a macromolecule has proven to be daunting; we shall return to this point below. Not surprisingly, approximations have had to be made. {\emph{Continuum solvent}} models retain the molecular detail of the macromolecule and treat the collective properties of the solvent implicitly, such as using a dielectric constant to describe the many-body electrical response of the solvent.  {\emph{Group additive}} approaches interpret the effect of solvent on the macromolecule by drawing upon more readily available hydration thermodynamics of small molecular groups comprising the macromolecule. The atomistic, continuum, and group additive models need not be mutually exclusive with elements of continuum models used in atomistic simulations and elements of group additive models used in continuum solvent simulations. 

Group additive approaches have had a long history in both experimental and theoretical studies of protein hydration, for example, see Refs.\ \citenum{kauzmann:59,tanford:62,tanford:apc70,Ooi:1987wt,Dill:1997tg}. In this approach the hydration free energies of small molecule analogs of groups comprising the macromolecule are additively combined to estimate the hydration free energy of the macromolecule. Typical efforts involve the intuitive idea of scaling the individual group contributions by its fractional exposure to the solvent. Lazaridis and Karplus attempted a more rigorous group contribution approach \cite{EEF1:protein99}, one that arose out of their clear recognition
that one must acknowledge the correlations that exist between binding energy distributions of different groups. They accounted for the correlation by developing a correction solely for the solvent exclusion effect.   The resulting effective energy function (EEF1) has proven to be influential in modeling macromolecular hydration; for example, EEF1 informs the hydration model in the popular ROSETTA protein structure prediction approach \cite{alford_rosetta_2017}. EEF1 has also informed the development of other models, in particular,  the ABSINTH \cite{absinth:jcc09} --- self-Assembly of Biomolecules Studied by an Implicit, Novel, and Tunable Hamiltonian --- model to be discussed below.
 
    
Treating the solvent as a dielectric in modeling hydration also has a long history. The most refined of these approaches solves the Poisson and Poisson-Boltzmann equations with full account of molecular structure, for example see Refs.\ \citenum{warwicker_calculation_1982,gilson_calculation_1988,yoon_computation_1992,honig_classical_1995}, but the resulting numerical system can become computationally demanding. Still and coworkers presented an approximation to solve the Poisson equation that they termed the generalized Born (GB) model \cite{still:jacs90}. This GB model is often supplemented with a model for the non-polar interactions, which are accounted by a surface area (SA)-dependent term \cite{qiu1997gb}. The resulting GB/SA model has also been popular in modeling the hydration of proteins. 

The GB/SA, EEF1, and ABSINTH approaches all represent important efforts to include the effect of the solvent albeit in an approximate way. However, for describing the thermodynamics of macromolecular hydration, 
a comparison of the predictions of these models versus all-atom results appears scarce, likely because of the challenges involved in calculating the hydration free energies of macromolecules.

Molecular quasi-chemical theory (QCT) \cite{lrp:apc02,lrp:book,lrp:cpms}  has enabled
entirely new computational studies ranging from biological macromolecule hydration to studies of water and aqueous ions involving rigorous first principle calculations.  In 2012,  in a first of its kind study, Weber and Asthagiri \cite{Weber:jctc12} tackled the challenging problem of calculating the hydration free energy of a protein, cytochrome C,  in an all-atom simulation. The developments following the 2012 study demonstrate that theoretical  refinements now make it possible 
\emph{to calculate the hydration thermodynamics of bio-macromolecules at the same resolution as for small molecules}. The emerging results have revealed fresh insights into the limitations of the additivity assumption \cite{paulaitis:corr10,tomar:bj2013,tomar:jpcb14,paulaitis:corr21}, explicated the unanticipated importance of long-range interactions in the role of denaturants \cite{tomar:gdmjcp18}, revealed the
critical role of solute-solvent attractive interactions in biomolecular
hydration \cite{tomar:jpcb16,asthagiri:gly15}, and most recently, showed that decades-old assumptions about hydrophobic hydration have a simple explanation in hydrophilic effects \cite{tomar:jpcl20}. 

Here we consider a range of polypeptides, including Gly$_{15}$, an archetype of intrinsically disordered peptides, and compare the hydration free energy based on molecular QCT against the values obtained using EEF1, ABSINTH, and GB/SA. We find that both EEF1 and ABSINTH are of limited utility in capturing the trends of hydration free energies predicted by molecular QCT. GB/SA does considerably better. On the basis of theory, we discuss the limitations of group additive models that have been a linch-pin in biophysical thinking. The insights here could spur refinements in models of protein solution thermodynamics.

\section{Theory} 

\begin{figure*}[h!]
\centering
\includegraphics[width=5.75in]{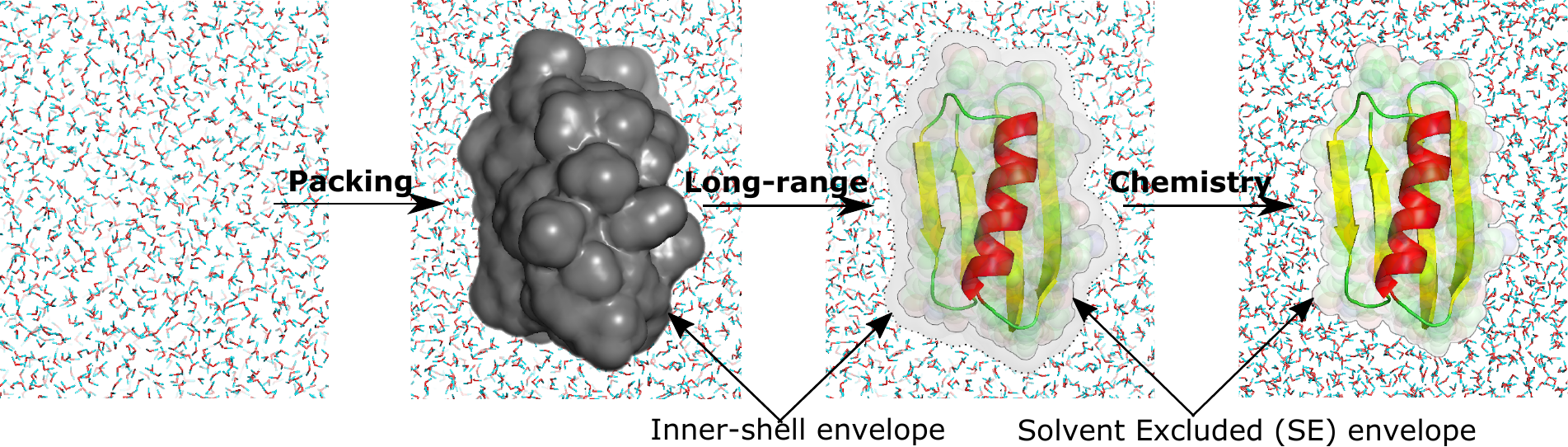}
\caption{Quasi-chemical organization of the excess chemical potential. The inner-shell  identifies the region enclosing the solute for which the solute-solvent binding energy distribution $P(\varepsilon|\phi)$ is accurately Gaussian. It approximately corresponds to the traditional first hydration shell of the solute. The free energy to create the cavity to accommodate the solute gives the packing (hydrophobic) contribution.  The chemistry contribution is zero for the solvent-excluded envelope.  The chemistry plus
long-range parts determine the hydrophilic contributions. Reprinted from Ref.~\citenum{tomar:jpcl20}, copyright (2020) American Chemical Society.}
\label{fg:fig1}
\end{figure*}

Eq.~\ref{eq:qct} provides the formal relation between the solute-solvent (water here) binding energy ($\varepsilon$) and the hydration free energy. However, a direct application of this relation is doomed to fail;  the binding energy distribution $P(\varepsilon)$ is usually an extreme value distribution and thus the high-$\epsilon$ tail of  this distribution is usually poorly sampled. We regularize this statistical problem by introducing an auxiliary field $\phi(r; \lambda)$ that moves the solvent away from the solute,
thereby tempering the solute-solvent binding energy.  The conditional distribution $P(\varepsilon|\phi)$ is better characterized than
$P(\varepsilon)$, and in calculations we adjust $\lambda$, the range of the field, to control the approximation of $P(\varepsilon|\phi)$ as a 
Gaussian distribution. With the introduction of the auxiliary field \cite{Weber:jctc12,tomar:bj2013,tomar:jpcb16} we get
\begin{eqnarray}
\beta \mu^{\mathrm{(ex)}} = 
\underbrace{- \ln p_0[\phi]}_{\rm packing} + 
\underbrace{\beta\mu^{\mathrm{(ex)}} [P(\varepsilon|\phi)]}_{\rm long-range} + 
\underbrace{\ln x_0[\phi]}_{\rm chemistry}~,
\label{eq:qc}
\end{eqnarray}
the quasi-chemical  organization of the potential distribution theorem \cite{lrp:book,lrp:cpms}.  The individual
contributions are functionals of the auxiliary field, as indicated, but the net excess potential is of course independent of $\phi$. 
Figure~\ref{fg:fig1} provides a schematic description of Eq.~\eqref{eq:qc}.

The auxiliary field in our calculations is based on a soft, repulsive potential \cite{chempath2009quasichemical,weber:jcp11}, specifically the WCA potential between water molecules. For this choice, we find that $\lambda \approx 5${\AA}  ensures that the conditional
binding energy distribution is accurately Gaussian, and we denote this range as $\lambda_\mathrm{G}$.  The largest value 
of $\lambda$ for which the chemistry contribution is negligible, labeled $\lambda_{\rm SE}$, has a special meaning. It bounds the domain excluded to the solvent.  We find $\lambda_{\rm SE} \approx 3${\AA}, unambiguously; see also Refs.~\citenum{tomar:jpcb16,asthagiri:gly15,asthagiri:jpcb20a}.  With this 
choice, Eq.~\eqref{eq:qc} can be rearranged as, 
\begin{eqnarray}
\beta \mu^{\mathrm{(ex)}}  =   \underbrace{-\ln p_0(\lambda_{\rm SE})}_{\rm solvent\, exclusion}  +  
\underbrace{\beta\mu^{\mathrm{(ex)}}[P(\varepsilon|\lambda_\mathrm{G})]}_{\rm long-range} + 
\underbrace{ \ln \left[x_0(\lambda_\mathrm{G})\left(\frac{p_0(\lambda_{\rm SE})}{p_0(\lambda_\mathrm{G})}\right)\right]}_{\rm revised\, chemistry}~.
\label{eq:qc1}
\end{eqnarray}
The several contributions are
identified by the range parameter. Thus, for example,
$x_0(\lambda_\mathrm{G})\equiv x_0[\phi(\lambda_\mathrm{G})]$. 
The revised chemistry term has the physical
meaning of the work done to move the solvent interface a distance
$\lambda_\mathrm{G}$ away from the 
volume excluded by the solute relative to the case when the
only role played by the solute is to exclude solvent up to $\lambda_{\rm
SE}$.  This term highlights the role of \emph{short-range} solute-solvent
attractive interactions on hydration. Interestingly, the range between
$\lambda_\mathrm{SE} = 3${\AA} and $\lambda_\mathrm{G} = 5${\AA}
corresponds to the first hydration shell for a methane
carbon \cite{asthagiri:jcp2008} and is an approximate descriptor of the
first hydration shell of groups containing nitrogen and oxygen heavy atoms. Eqs.~\ref{eq:qc} and \ref{eq:qc1} are rigorous and complete. 

We use the above development to examine the EEF1, ABSINTH, and GB/SA implicit-solvent models.

\subsection{EEF1}

Imagine decomposing the solute into a collection of well-defined groups, such as for example, $\mathrm{CH_3-}$, $\mathrm{-C(O)N(H)-}$, etc. We index the group by $i$. The solute-water binding energy distribution of each group, in the absence of all the other groups, is $P_i(\varepsilon)$. If in composing the solute the groups can be treated as independent of each other, then the binding energy distribution of the solute is $P(\varepsilon) = \Pi_i\, P_i(\varepsilon)$; thus $\mu^{\mathrm{(ex)}} = \sum_i \mu^{\mathrm{(ex)}}_i$, where $\mu^{\mathrm{(ex)}}_i$ is the hydration free energy of group $i$. 

The EEF1 \cite{EEF1:protein99} recognizes that $P(\varepsilon) \neq \Pi_i\, P_i(\varepsilon)$ because of the presence of correlations between groups in the physical solute.  For example, two isolated cavities that exclude water will do so to a lesser extent when they are bonded.  In EEF1 the tacit assumption is that this solvent exclusion effect is a major contributor to the correlations and the effect is accounted for as
\begin{equation}
      \mu^{\mathrm{(ex)}}_{\mathrm{EEF1}} = \sum_i \mu^{\mathrm{(ex)}}_i - \sum_{i} \sum_{j \neq i} f_{i} (r_{ij}) V_{j} \, ,
\label{eq:eef1}
\end{equation} 
where the function $f_i$, the solvation free energy density,  is a Gaussian function of the distance $r_{ij}$  between groups $i$ and $j$, and $V_j$ is the volume of the group $j$.  $\mu^{\mathrm{(ex)}}_i$, the isolated group hydration free energy, is based on the hydration of reference groups. Equation \ref{eq:eef1} is implemented in CHARMM \cite{brooks2009charmm} by switching on the EEF1 force field. From the perspective of molecular QCT, we can recognize that EEF1 ignores correlations in the chemistry and long-range hydrophilic contributions and uses an approximate, albeit convenient, model for short-range solvent exclusion. 

\subsection{ABSINTH}

In the ABSINTH \cite{absinth:jcc09} model, like EEF1, one decomposes the polypeptide into smaller groups with known hydration free energies. However, the groups considered in ABSINTH are larger than those considered in EEF1, and a different functional form is used for treating the solvent exclusion effect. Additionally, ABSINTH also incorporates the role of intra-molecular charge-charge interactions to the electrostatic energy of a solute. The hydration free energy within ABSINTH is given as  
\begin{equation}
    \mu^{\mathrm{(ex)}}_{\mathrm{ABSINTH}}  = \sum_i \zeta_{i} \, \mu^{\mathrm{(ex)}}_i + W_{el} \Big |_{78.2} - W_{el} \Big |_{1.0},
\label{eq:abs} \, 
\end{equation}
where $\mu^{\mathrm{(ex)}}_i$ are the hydration free energies of reference groups in the absence of inter-group correlations, $\zeta_i$ are the factors that account for solvent exclusion, and the last two terms on the right hand side account for the 
change in electrostatic energy (from charge-charge interactions) in moving the solute from vacuum (dielectric 1.0) to water (dielectric 78.2).  Note that charge self-interaction --- the Born term for an isolated charge --- is not included in Eq.~\ref{eq:abs}. 
The first term on the right hand side of Eq.~\ref{eq:abs} is called the Direct Mean Field Interaction (DMFI) between the solute and the solvent.

Although our attempts to derive the above approximation from Eq.~\ref{eq:qc1} have not proven successful,  we can discern that ABSINTH attempts to model the correlations in both the hydrophilic and hydrophobic contributions
by a local factor $\zeta_i$. Additionally, the model attempts to incorporate collective properties of hydration by accounting for the role of the solvent dielectric in modulating
charge-charge interactions.

\subsection{GB/SA}

Molecular QCT (Eq.~\ref{eq:qc1}) naturally separates the roles of attractive and repulsive contributions to hydration. 
The GB/SA \cite{still:jacs90,qiu1997gb} decomposition of the hydration free energy is similar in spirit. 
The contribution arising due to the attractive solute-water interactions is modeled by treating the solvent as a continuum dielectric and by using the so-called Coulomb field approximation \cite{bashford_generalized_2000}. For inter-charge interactions, an interpolation approach is used to go from 
an effective Born radius at short range to the distance itself at long-range. In the GB formalism, changes in free energy due to charge-charge interactions and the change in the self-energy of a charge (the Born term) are both included. 
The role of attractive interactions arising from non-electrostatic effects (a negative contribution to the free energy) and the solvent exclusion contribution (a positive contribution to the free energy) is lumped in the surface area term. Note that in the GB-approach, collective properties of hydrophilic hydration are naturally included, albeit within a continuum dielectric framework. 
The SA-terms can also be seen as a model for the collective properties of non-polar hydration. 

\section{Methods} 

Results from molecular QCT are taken from published data appearing over several papers --- the hydration free energy change for Gly$_n$ ($n=1\ldots 5$) \cite{tomar:bj2013}, 
context (position) dependence of the hydration of isoleucine in GGIGG or IGGGG chains \cite{tomar:jpcb14}, 
the hydration free energy of Gly$_{15}$ \cite{asthagiri:gly15} in various conformations, and the hydration 
free energies for deca-alanine helices and coils \cite{tomar:jpcb16}. New results are obtained for capped alanine peptide and a fuller account of the approach is in the SI. 

The hydration free energies using ABSINTH are calculated using its implementation in the CAMPARI (v4) \cite{campari:2009} simulation engine. The {\sc PDBANALYZE} feature is used (once with a dielectric of 78.2, and a second time with a dielectric of 1.0) to calculate the hydration free energies of structures.

To study the $\phi-\psi$ dependence of $\mu^{\mathrm{(ex)}}$ for capped alanine, we used 
CAMPARI \cite {campari:2009} to sample structures.  (This is done to exploit the superior sampling of configurations in $\phi-\psi$ space by the ABSINTH Monte Carlo routine. Even for the simple molecule considered here, there are basins in $\phi-\psi$ space that are excluded from a naive molecular dynamics sampling using GB/SA. Incomplete sampling from molecular dynamics simulations is studied more comprehensively in other works \cite{huber1994local, darve2001calculating, babin2009adaptively}.) The alanine residue capped with ACE and NME groups is placed in a spherical water droplet of radius 100~\AA. The temperature of the system is set to 298.15 K. After 50 million sweeps of equilibration, structures are saved every 2000 sweeps for the next 100 million sweeps. 10 such independent simulations are performed to obtain 500,000 structures of capped alanine. The structures generated from ABSINTH are used to calculate  $\mu^{\mathrm{(ex)}}$ from all 3 implicit solvent models.  The $\mu^{\mathrm{(ex)}}$ values are binned in ($2.5^\circ \, \times \, 2.5^\circ$) grids. 
Since the variance in $\mu^{\mathrm{(ex)}}$ within a grid is negligible, we simply report the naive average of the $\mu^{\mathrm{(ex)}}$ per bin. (Note that the multi-state generalization of the potential distribution theorem provides a rigorous way to conformationally average thermal quantities, for example Ref.\ \citenum{asthagiri:gly15}, but that is not needed here. Also see the SI.)

All of our GB/SA calculations are performed using NAMD \cite{namd:2020}. The {\sc GBIS} keyword is used to turn-on 
the GB calculations. The cut-off  distance is set to 14 \AA\ (alphaCutoff).  For the SA part, 
the ``surface tension" parameter is set to 0.00725 kcal/{\AA}$^2$, a value that we find empirically to best match the slope of hydration free energies of Gly$_n$ obtained using molecular QCT.  
$\mu^{\mathrm{(ex)}}$ is obtained by a simple subtraction of the configurational ``potential" energy of the solute in vacuum, from the configurational ``potential" energy of the solute in the continuum
solvent. 

In all-atom simulations, the pressure is always 1~atm and the temperature is either 298.15~K or 300~K. In implicit solvent models, an explicit pressure does not enter, but the conditions (model parameters) are understood to correspond to 1~atm pressure. 

\section{Results}

\subsection{Hydration of linear Gly$_n$ peptides}

\begin{figure*}[ht]
\begin{center}
 \includegraphics[width=0.975\textwidth]{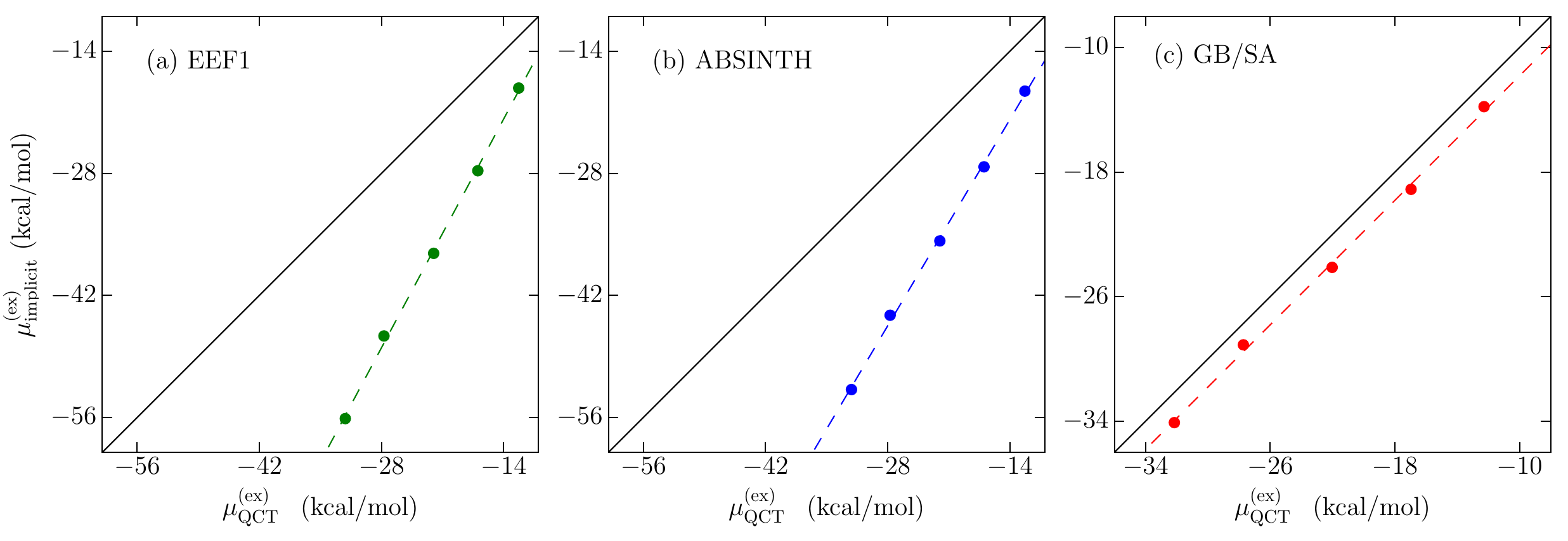}
\end{center}
\caption{$\mu^{\mathrm{(ex)}}$ from implicit solvent models versus molecular QCT results for straight chain Gly$_{n}$ ($n=1\ldots 5$). All calculations are at 298.15 K. Symbols are simulation data, and dashed  best fit line is 
a guide to the eye. The molecular QCT results are obtained with permission from figure 3 of Ref. \citenum{tomar:bj2013}, copyright Biophysical Journal.}
\label{fg:gly_chains}
\end{figure*}

Fig.~\ref{fg:gly_chains} compares $\mu^{\mathrm{(ex)}}$ from implicit solvent models versus the results from all-atom molecular QCT for linear Gly$_n$, $n = 1\ldots 5$.  With a slight tuning of the surface tension parameter the results from the GB/SA model (Fig.~\ref{fg:gly_chains}(c)) can be made nearly parallel to the molecular QCT results.  (For all our calculations in this work, we use this value for the surface tension parameter.)  Please note that $\mu^{\mathrm{(ex)}}$ decreases linearly with increasing $n$ for all the models. Fig.~\ref{fg:gly_chains} thus makes it clear 
that relative to molecular QCT both EEF1 and ABSINTH 
greatly over-predict $\mu^{\mathrm{(ex)}}$ for higher $n$. 

\subsection{Context dependence of hydration in isoleucine-glycine peptides}

Polyampholytes are understood to be a more complex system to study than polyelectrolytes, because of the many possible 
residue combinations \cite{higgs1991theory,das:pnas2013,lin2016sequence, dinic2021polyampholyte}. In light of this understanding, the sensitivity of implicit solvent formalisms to the spatial position of residues needs to be tested. In this work, we start with a simple chain of GGGGG pentapeptide and replace one of the G along the chain with an isoleucine (I). 
Specifically, we consider IGGGG and GGIGG. The difference in hydration free energy for the two chains from all three implicit models are compared to molecular QCT results from Ref. \citenum{tomar:jpcb14}. Table~\ref{tb:solv_free_en} shows an under-appreciation (almost by an order of magnitude) of the distinctiveness of the two chains by all of the implicit solvent models considered. 

\ctable[
	mincapwidth=\textwidth,
	caption={Context dependence of hydrophobic hydration in isoleucine-glycine chains.  $\mu^{\mathrm{(ex)}}$ of GGIGG or IGGGG is shown relative to GGGGG. All calculations are at 298.15 K. $\Delta = (\mu^{\mathrm{(ex)}}_\mathrm{GGIGG} - \mu^{\mathrm{(ex)}}_\mathrm{IGGGG})/\mu^{\mathrm{(ex)}}_\mathrm{GGGGG}$. 
		All energies are in kcal/mol.},
	label=tb:solv_free_en,
	pos=h,
	captionskip=-1.5ex
	]	
{c c c c c }
{
\tnote{The molecular QCT data is obtained with permission from table 2 of Ref. \citenum{tomar:jpcb14}, copyright Americal Chemical Society. }
}
{
\FL
Model Type                &   GGGGG       &  IGGGG &   GGIGG &  $\% \Delta$ \ML
ABSINTH                   &    $-52.8$            &  $+3.4$      &    $+3.5$     &  0.2                  \NN[-1ex]
EEF1                         &     $-56.1$           &  $+5.6$       &     $+5.9$    &  0.4                 \NN[-1ex]
GB/SA                       &     $-34.1$    &  $+2.7$          &     $+2.9$         &  0.4                \NN[-1ex]
Explicit Water             &     $-32.2$     &   $+2.1$       &      $+2.9$        &  2.5           \LL
}

\subsection{Hydration in the collapse transition of Gly$_{15}$}

\begin{figure*}[ht]
\begin{center}
  \includegraphics[width=0.975\textwidth]{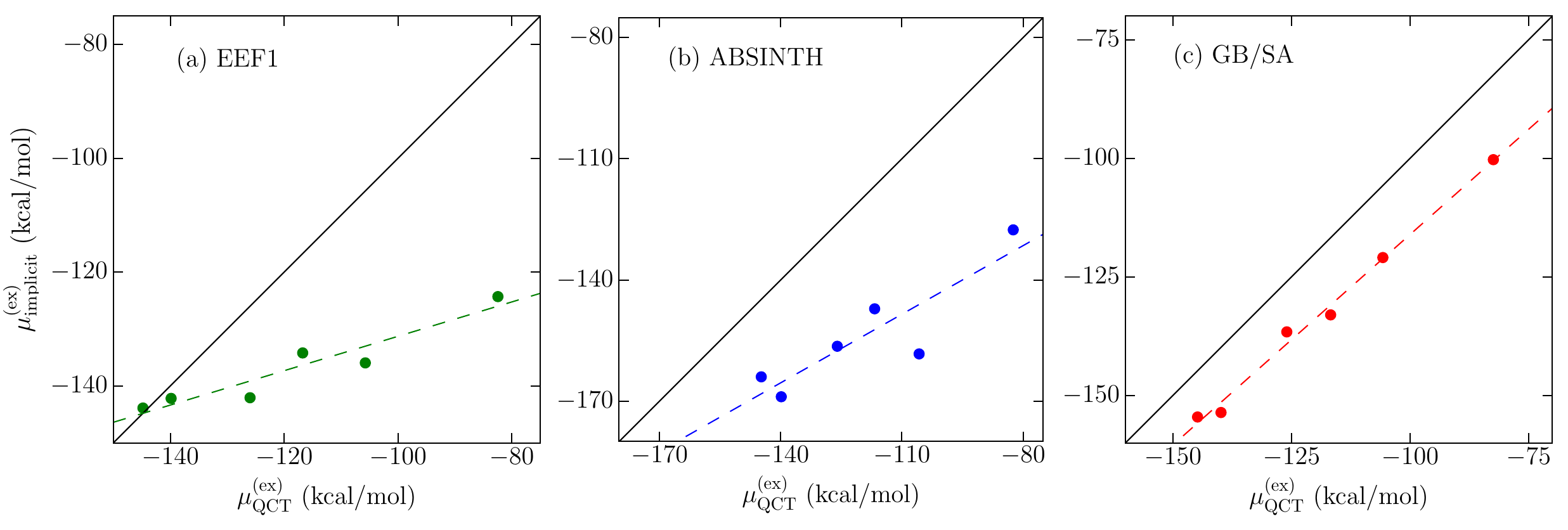}
\end{center}
\caption{$\mu^{\mathrm{(ex)}}$ from implicit solvent models versus molecular QCT results for Gly$_{15}$ hydration in various conformations. All calculations are at 300 K.  Rest as in Fig.~\ref{fg:gly_chains}. Molecular QCT results are obtained with permission from figure 2 of Ref. \citenum{asthagiri:gly15}, copyright American Chemical Society.}
\label{fg:poly_gly}
\end{figure*}

Studies on oligoglycines have indicated their preference for collapsed conformations in solution \cite{kiefhaber:pnas06, Teufel:2011jmb,Tran:2008bk,Hu:2010b}.  Our earlier study \cite{asthagiri:gly15} on Gly$_{15}$ shows that hydration tends to expand the polypeptide and the collapse is driven by intra-molecular interactions over-coming opposing hydration effects. 
Thus the difference in hydration free energies between collapsed and expanded states is a major factor in their relative stability. Given this sensitivity, the physics of oligoglycine collapse is an interesting problem to look at from the perspective of implicit solvent models. Fig.~\ref{fg:poly_gly}(c) shows that predictions based on
GB/SA follow the same relative ordering of $\mu^{\mathrm{(ex)}}$
as those based on molecular QCT.  However, the relative ordering breaks down for both EEF1 and ABSINTH. Quantitatively, the difference in $\mu^{\mathrm{(ex)}}$ between the most collapsed and expanded structures calculated based on GB/SA ($ \approx 54$ \, kcal/mol) is closest to the molecular QCT prediction ($ \approx 62$ \, kcal/mol).
For ABSINTH this difference is $ \approx 46$  kcal/mol and for EEF1 it is $\approx 20$  kcal/mol.
Clearly, EEF1 predicts the least amount of deviation in  $\mu^{\mathrm{(ex)}}$ between the most collapsed and most expanded structures. 

\subsection{Hydration in helix-coil transition of deca-alanine}

Tomar et al. \cite{tomar:jpcb16} have examined the balance of hydrophobic and hydrophilic effects in the helix coil transition of deca-alanine. Molecular QCT results show that coils are better hydrated because of the dominance of hydrophilic effects, and that the coil-to-helix transition is driven by the net intra-molecular interactions within the solute. 
Testing implicit solvent models on this ubiquitous problem in biochemistry can prove illuminating. 

\begin{figure*}[ht]
\begin{center}
 \includegraphics[width=0.975\textwidth]{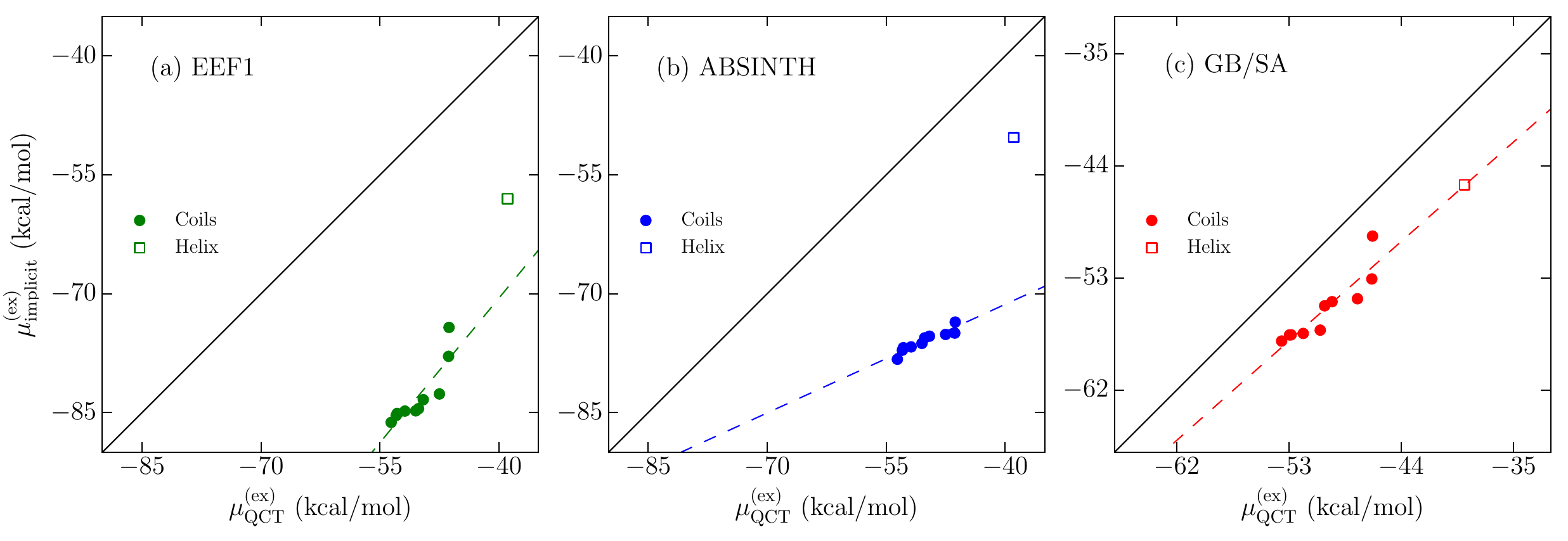}
\end{center}
\caption{$\mu^{\mathrm{(ex)}}$ from implicit solvent models versus molecular QCT for the hydration of several coil conformers and a helix conformer of deca-alanine. All calculations are at 298.15 K.  Rest as in Fig.~\ref{fg:gly_chains}. 
Molecular QCT results are obtained with permission from figure 2 of Ref. \citenum{tomar:jpcb16}, copyright American Chemical Society.}
\label{fg:ala_10}
\end{figure*}

Fig.~\ref{fg:ala_10}  compares the prediction of $\mu^{\mathrm{(ex)}}$ for several coil conformers and a helix conformer from different methods. Once again GB/SA does better in capturing the relative trends and magnitudes of  $\mu^{\mathrm{(ex)}}$.  EEF1 and ABSINTH both predict a greater
gap in  $\mu^{\mathrm{(ex)}}$ between coil conformers and the helix. The relatively high difference in ABSINTH arises due to the large change in $W_{el}$ for the helix in vacuum versus that in water.

\subsection{$\phi$-$\psi$ $\mu^{\mathrm{(ex)}}$ landscape of capped alanine}

\begin{figure*}[ht]
\begin{center}
  \includegraphics[width=0.99\textwidth]{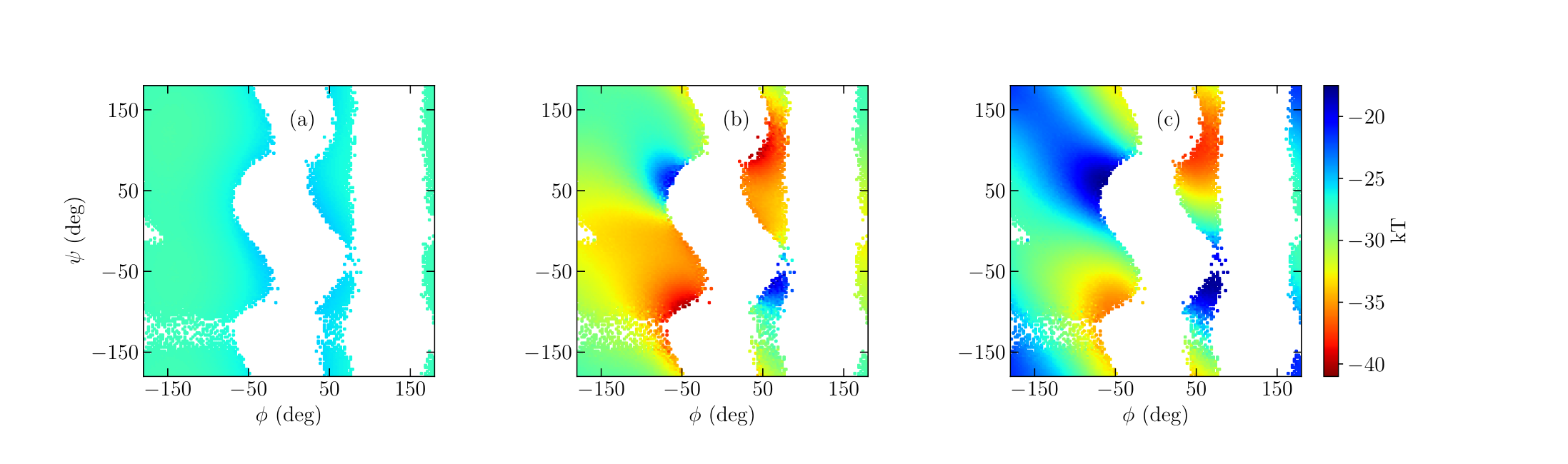}  
\end{center}
\caption{Distribution of $\mu^{\mathrm{(ex)}}$ in the $\phi$-$\psi$ space for capped (ACE-ALA-NME) alanine using the
implicit models (a) EEF1, (b) ABSINTH, and (c) GB/SA. All calculations are at 298.15 K.}
\label{fg:ala_cmap}
\end{figure*}

Earlier, Choi and Pappu \cite{choi2018experimentally} obtained experimentally derived and computationally optimized 
$\phi$-$\psi$ landscapes \cite{ramachandran1968conformation}  for all the amino acids  and on this basis they have sought to 
improve the ABSINTH Hamiltonian \cite{choi2019improvements}. The improvement takes the form of correction terms 
to the dihedral potential.  The potential of mean force (PMF) in the $\phi$-$\psi$ space is an aggregate effect of 
all of the terms in the Hamiltonian.  Here we seek to analyze the role of hydration alone in the
$\phi-\psi$ distribution (Fig.~\ref{fg:ala_cmap}).

Qualitatively, the most striking difference amongst the hydration landscapes in Fig.~\ref{fg:ala_cmap} is their diversity. EEF1 (Fig.~\ref{fg:ala_cmap}(a)) in comparison to both ABSINTH (Fig.~\ref{fg:ala_cmap}(b)) and GB/SA (Fig.~\ref{fg:ala_cmap}(c)) is 
rather featureless.  ABSINTH shows a richer diversity in $\mu^{\mathrm{(ex)}}$ than EEF1; however, relative to GB/SA
it has larger areas of uniformity. The higher diversity of $\mu^{\mathrm{(ex)}}$ from ABSINTH compared to EEF1 can be
traced to the electrostatic term ($W_{el}$). (See also the SI.) Our results reaffirm the intuition that a more explicit incorporation of electrostatics leads to greater diversity in $\mu^{\mathrm{(ex)}}$ \cite{absinth:jcc09}.   

For a quantitative comparison with molecular QCT, we select some definite structures. Computationally optimized PMFs from experimental structures of capped alanine show 4 major basins \cite{choi2018experimentally}. We pick the most stable ABSINTH structure in each of these regions. Additionally, we also pick 2 structures where ABSINTH and GB/SA hydration free energies differ the most. Results from all 6 structures are tabulated in \ref{tb:ala_exp} (from the least stable to the most stable as predicted by molecular QCT). Quantitatively, GB/SA is in better agreement with molecular QCT than ABSINTH. Qualitatively, GB/SA also better captures the ordering of the conformers than ABSINTH.

\ctable[
	mincapwidth=\textwidth,
	caption={Comparison of hydration free energies at 298.15 K for select structures of capped alanine. All energies are in kcal/mol.},
	label=tb:ala_exp,
	pos=h,
	captionskip=-1.5ex
	]	
{c r r c c c }
{}
{
\FL
Structure   &  $\phi$ (deg)   & $\psi$ (deg)  & ABSINTH  &  GB/SA  &  Molecular QCT \ML
1   &  $-146.86$     &  $151.28$   &  $-16.7$    & $-13.5$  &    $-12.2$  \NN[-1ex]
2   &   $-65.96$     &   $13.71$   & $-20.5$     & $-14.3$  &    $-14.2$ \NN[-1ex]
3   &    $50.90$     &   $3.54$    &  $-21.3$    & $-15.7$  &    $-16.2$ \NN[-1ex]
4   &   $-64.14$     &  $145.12$   & $-17.0$     & $-17.0$  &    $-17.3$  \NN[-1ex]
5   &   $-64.19$     &  $-36.09$   &  $-20.8$    &  $-18.6$ &    $-19.5$ \NN[-1ex]
6   &    $54.42$     &   $29.90$   &  $-20.5$    & $-18.3$  &    $-19.8$ \LL 

}

The above results show that an inadequate treatment of hydration can be masked by adjusting some other aspect of the energy function, limiting the utility of the energy function to conditions for which the function has been tested and tuned.

\section{Discussion}

The results above indicate that even an approximate description of the collective properties of hydration seems to be better than using an additive model of hydration. With some exceptions, for all the cases studied here 
the implicit solvent models yield $\mu^{\mathrm{(ex)}}$ values that are more negative compared to the all-atom results based on molecular QCT.  

The simplicity of the GB/SA approach allows one to readily identify its potential deficiencies --- the use of a dielectric (at short range), the Coulomb field approximation for the electric displacement, and a surface area model that lumps non-polar and hydrophobic contributions. It is well known from scaled-particle theory, that a surface area description for modeling 
hydrophobic hydration is problematic for small length scales \cite{Ashbaugh:rmp}.  The limitations of a dielectric description 
have also been extensively studied in the literature. Nevertheless, relative to EEF1 and ABSINTH, the GB/SA approach comes closest in describing the results from molecular QCT. 

The limitations of EEF1 and ABSINTH, especially in calculating thermal quantities such as free energies and entropies relate to a subtler issue, namely the assumption of group additivity.  In an early study, Roseman \cite{Roseman:jmb88} had suggested the limitations of additivity by comparing theoretically derived group-transfer free energies based on structure-additivity relation  versus experimental results. Free energy calculations on helices of different lengths \cite{helms:2005fw}, of pairs of blocked amino acids \cite{Boresch:jpcb09}, and of blocked neutral side chains \cite{Boresch:bj13} reached similar conclusions.  The 2012 study by Weber and Asthagiri \cite{Weber:jctc12} hinted at the shortcomings of the EEF1 model, and subsequent studies brought more physical clarity to the limitations of 
of group additivity for thermal quantities \cite{paulaitis:corr10,tomar:bj2013,tomar:jpcb14,paulaitis:corr21}. More recently
we showed that results attributed to hydrophobicity by group additive methods in fact arise from hydrophilic effects \cite{tomar:jpcl20}. 

Here we bring the clarity afforded by molecular theory to understand the characteristic discrepancies in EEF1 and ABSINTH results from two different perspectives. 

\subsection{Correlation effects between different groups}

\begin{figure*}[ht]
\begin{center}
  \includegraphics[width=0.49\textwidth]{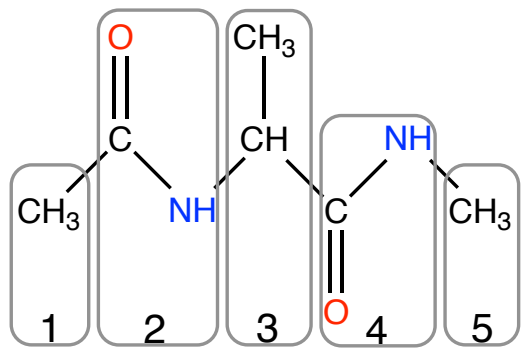}
\end{center}
\caption{Division of a capped alanine into five groups for our analysis of correlation effects}
\label{fg:ala_group}
\end{figure*}

Consider the capped alanine molecule and its division into groups shown below (Fig.~\ref{fg:ala_group}). 
Let us evaluate the packing, chemistry, and long-range contributions to hydration from a group additive approach. 
Recall that the chemistry contribution is the negative of the work done to move water away from the solute to beyond the defined inner shell. We seek this for a defined group (Fig.~\ref{fg:ala_group}), with the rest of the solute fully interacting with the solvent. In a similar way, we obtain packing contributions for just the group. 
The long-range contributions are obtained solely on the basis of individual group binding energy distribution with the solvent. 

\ctable[
	mincapwidth=\textwidth,
	caption={Group-wise decomposition of the quasi-components (Eq.~\ref{eq:qc1}) in the hydration of capped alanine Fig.~\ref{fg:ala_group}. The non-additive values are based on treating the solute as a single entity. All the energies are in kcal/mol.},
	label=tb:ala_group,
	pos=h,
	captionskip=-1.5ex
	]	
{c r r r }
{}
{
\FL
Group &  Revised chemistry & Long-range  & Packing  \ML
1                          &  2.4   &  -2.1 & 2.3  \NN[-1ex]
2                          &  -0.4  &  -4.4 & 4.2 \NN[-1ex]
3                          &  3.1   &   0.1 & 3.4  \NN[-1ex]
4                          &  -1.7  &  -4.3 & 4.2 \NN[-1ex]
5                          &  2.0   &   -0.2 & 2.3 \NN[-1ex]
Sum                        & 5.4   & -10.9 &  16.4 \NN[-1ex]
Non-additive & $\mathbf{-13.6}$ & $\mathbf{-9.6}$ & $\mathbf{11.0}$ \LL
}

Consider the non-polar groups 1, 3, and 5 (Fig.~\ref{fg:ala_group}). Table~\ref{tb:ala_group} shows that the revised chemistry contribution is positive, just as was found for an isolated methane in water \cite{asthagiri:jcp2008,tomar:jpcl20}. This seemingly suggests that, as found for methane, water is pushed into contact with these non-polar groups. 
For the polar peptide groups, the revised chemistry contribution is negative. The group-additive estimate is net positive, and \textit{dramatically in error} relative to the large negative value obtained by treating the solute as a single entity. 
To understand the physical reason for this failing, recognize that less work will be needed to evacuate the inner shell for, say, group 2 (Fig.~\ref{fg:ala_group}) when the inner shell around, say, group 1 is already evacuated. This cooperativity of hydration is the reason why additivity fails.  

For the same reasons noted above,  the packing contribution treating the solute as a collection of groups is more positive than treating the solute as a single entity; see also Ref.\ \citenum{tomar:bj2013}. For the long-range contributions, as expected on physical grounds the cooperativity is weaker \cite{tomar:bj2013, paulaitis:corr21,paulaitis:corr10}. and the deviation between a group additive and non-additive descriptions is small (relative to the large deviations seen for chemistry and packing contributions). However, the cooperativity does not go away entirely even for the long-range contributions. 


 Group additive models were at the heart of ideas implicating hydrophobic hydration as a dominant force in protein folding and assembly \cite{kauzmann:59,tanford:62,dill:1990ww} and also find  expression in numerous other instances in biophysics.  Our recent study \cite{tomar:jpcl20}  and this work exposes the physical and conceptual limitations of group additivity, emphasizing that there is a distinct possibility of deriving incorrect physical conclusions based on such models.

\subsection{Alternative perspective of correlation effects}

Consider the hydration of a dimer wherein the monomers are a distance $r$ apart.  The potential distribution approach readily allows one to relate the chemical potential of the dimer to the excess chemical potential of the monomer as \cite{widom:jcp63,widom:jpc82,asthagiri:jcp2008}
\begin{equation}
      \beta \mu_{2}^{\mathrm{(ex)}}(r)  =  2 \beta \mu_{1}^{\mathrm{(ex)}} -  \ln\left[ y(r)\right] \, , 
\label{eq:widom_2}
\end{equation}
where $y(r) = g(r) \exp[u(r) / k_{\rm B}T]$, here $g(r)$ is the usual pair-correlation function and $u(r)$ is the pair potential. $\ln [y(r)]$ can be interpreted as the work necessary to bring two monomers to the bond length. For a general $n$-mer solute, we have\cite{stell:fpe92} 
\begin{equation}
      \beta \mu_{n}^{\mathrm{(ex)}}(\mathbf{r_1,\ldots,r_n})  =  \sum\limits_{j=1}^n \beta \mu_{j}^{\mathrm{(ex)}} -  \ln\left[ y(\mathbf{r_1,\ldots,r_n})\right] \, . 
\label{eq:widom_n}
\end{equation}
Assuming an $n$-mer chain of identical monomers with monomers uniformly spaced r apart, and further assuming that the pair correlation between each monomer pair is independent of the chain length, we have\cite{ghonasgi1993henry}
\begin{equation}
      \beta \mu_{n}^{\mathrm{(ex)}} \, =  n \beta \mu_{1}^{\mathrm{(ex)}} - (n-1) \ln \left[ y(r) \right]
\label{eq:cg} \, .
\end{equation}
Analyzing the straight-chain Gly$_n$ ($n=1\ldots 5$) hydration using Eq.~\ref{eq:cg} gives $-k_{\rm B}T \ln [y(r)]$ as follows: explicit water ($ \approx 7.1 $ kcal/mol), EEF1 ($\approx 8.7 $ kcal/mol), ABSINTH ($ \approx 10.1 $ kcal/mol), and GB/SA ($ \approx 8.9 $ kcal/mol).  Evidently, all the implicit solvent models predict a weaker cooperativity (more positive $-\ln y(r)$) of hydration  than the explicit solvent model.

 In EEF1 cooperativity in short-range hydrophobic hydration is entirely ignored and a correction developed only for the packing contribution. In ABSINTH and earlier surface area based models  \cite{Ooi:1987wt}, the effect of cooperativity in both short-range hydrophilic and hydrophobic contributions is modeled by scaling the reference group hydration free energies. While this can lead to better numerical results in some cases, the molecular QCT results and Eqs.~\ref{eq:widom_2}, ~\ref{eq:widom_n}, ~\ref{eq:cg}  highlight the limitations of such models. (See also Ref.\ \citenum{tomar:jpcb14}.)

\section{Conclusions}

For simple poly-glycine and poly-alanine peptides, a qualitative and quantitative comparison of predictions based on EEF1, ABSINTH, and GB/SA models shows that GB/SA is best able to capture the explicit water results based on molecular QCT (barring an offset, which is not of concern for our studies). ABSINTH and EEF1 
do not provide a credible description of the thermodynamics of hydration for these systems. 

Examining the role of hydration in the $\phi$-$\psi$ distribution for a blocked alanine peptide reveals the critical role of electrostatics in discriminating structures in different basins. In this regard, by treating the collective properties of hydration albeit with a simple dielectric constant, GB/SA proves better than either ABSINTH or EEF1. However, with some exceptions, all three implicit solvent models predict a more negative (more favorable) hydration than the results based on a molecular description of water. This result is ultimately related to the inadequate description of cooperativity (of hydration) in implicit solvent models. Our results for the $\phi$-$\psi$ landscape of capped alanine also indicate the foundational relevance of refining individual components of an implicit solvent model. Attempts to tune the dihedral terms of the energy function, for example, can at best mask the deficiencies of a different kind of physics, in this case in describing hydration. These comments also apply to how physics-based scoring functions are designed and developed for predicting protein structures or studying drug-protein interactions. 

The limitations of the implicit solvent models notwithstanding, the development here suggests possible ways to improve implicit solvent models, which after all remain important in biophysical studies. For example, it could be helpful to better tune the SA model in GB/SA to predict hydration of more complex species. One can also envision
using correlations obtained from explicit solvent hydration data to inform multibody exclusion models in ABSINTH or EEF1.  A critical
examination of configurational sampling from different implicit solvent models could also prove helpful in guiding future developments of such models. 

\section{Acknowledgements}

We gratefully acknowledge the Robert A. Welch foundation (Grant No. C-1241) for their financial support. 
This research used resources of the National Energy Research Scientific Computing Center, which is supported by the Office of Science of the U.S. Department of Energy under Contract \# DE-AC02-05CH11231. 
 We also gratefully acknowledge the Texas Advanced Computing Center (TACC) at The University of Texas at Austin (URL: http://www.tacc.utexas.edu) for providing HPC resources. 


\providecommand{\latin}[1]{#1}
\makeatletter
\providecommand{\doi}
  {\begingroup\let\do\@makeother\dospecials
  \catcode`\{=1 \catcode`\}=2 \doi@aux}
\providecommand{\doi@aux}[1]{\endgroup\texttt{#1}}
\makeatother
\providecommand*\mcitethebibliography{\thebibliography}
\csname @ifundefined\endcsname{endmcitethebibliography}
  {\let\endmcitethebibliography\endthebibliography}{}

\end{document}